\documentclass[amsfonts,showpacs,tightenlines,aps,12pt,floatfix]{revtex4}
\usepackage{bm}
\usepackage{epsfig}
\begin{document}


\title{Constraints on the $I=1$ hadronic $\tau$ decay and 
$e^+e^-\rightarrow hadrons$ data sets and implications 
for $(g-2)_\mu$}
\author{Kim Maltman}
\email[]{kmaltman@yorku.ca}
\affiliation{CSSM, Univ. of Adelaide, Adelaide, SA 5005 AUSTRALIA}
\altaffiliation{Permanent address: 
Department of Mathematics and Statistics, York University, 
4700 Keele St., Toronto, ON CANADA M3J 1P3}

\date{\today}

\begin{abstract}
Sum rule tests are performed on the spectral data for
(i) flavor $ud$ vector-current-induced hadronic $\tau$ decays and 
(ii) $e^+e^-$ hadroproduction, in the region below 
$s\sim 3-4\ {\rm GeV}^2$, where discrepancies exist 
between the isospin-breaking-corrected 
charged and neutral current $I=1$ spectral functions.
The $\tau$ data is found to be compatible with expectations based
on high-scale $\alpha_s(M_Z)$ determinations, while
the electroproduction data displays two problems.
The results
favor determinations of the leading order hadronic
contribution to $(g-2)_\mu$ which incorporate
hadronic $\tau$ decay data over those employing
electroproduction data only, and hence a reduced discrepancy
between experiment and the Standard Model prediction for $(g-2)_\mu$.

\end{abstract}

\pacs{13.35.Bv,13.40.Gp,13.66.Bc,13.35.Dx}

\maketitle

\section{\label{intro}Introduction}
In the Standard Model (SM), the largest of the 
non-purely-leptonic contributions 
to the anomalous magnetic moment of the muon, $a_\mu \equiv (g-2)_\mu /2$,
is that due to the leading order (LO) hadronic vacuum polarization (VP),
$\left[ a_\mu\right]_{had}^{LO}$. $a_\mu$ is currently known to
$0.5$ ppm~\cite{bnlgminus2}, with a proposal to reduce
this to $0.2$ ppm in the near future~\cite{newbnlproposal}. 
The $0.5$ ppm uncertainty represents $< 1\%$ of 
$\left[ a_\mu\right]_{had}^{LO}$, making an accurate determination of 
$\left[ a_\mu\right]_{had}^{LO}$ crucial to the study of
possible non-SM contributions to $a_\mu$.

$\left[ a_\mu\right]_{had}^{LO}$ is related to 
$\sigma\left[ e^+e^-\rightarrow hadrons\right]$ by the dispersion 
representation~\cite{gdr69}
\begin{equation}
\left[ a_\mu\right]_{had}^{LO}\, =\, {\frac{\alpha^2_{EM}(0)}{3\pi^2}}\,
\int_{4m_\pi^2}^\infty ds {\frac{K(s)}{s}}\, R(s)\ ,
\label{gm2weight}\end{equation}
where the form of $K(s)$ is well-known, 
and $R(s)$ is the ``bare'' $e^+e^-\rightarrow hadrons$ to
$e^+e^-\rightarrow\mu^+\mu^-$ cross-section ratio. With recent 
electroproduction data, the uncertainty on $\left[ a_\mu\right]_{had}^{LO}$ 
from Eq.~(\ref{gm2weight}) is comparable to the experimental
error on $a_\mu$, and dominates the uncertainty in the SM prediction for 
$a_\mu$~\cite{dehz02,dehz03,hmnt04,hockerichep04,otherdispersiveamu}.
Since CVC relates the $I=1$ 
electromagnetic (EM) spectral function to the 
charged current vector spectral function measured
in $\tau^-\rightarrow\nu_\tau +$ {\it non-strange hadrons}, 
hadronic $\tau$ decay data~\cite{aleph97,opal99,cleo00} can,
in principle, be used to improve the determination of 
$\left[ a_\mu\right]_{had}^{LO}$~\cite{adh98,dh98}. At the $<1\%$
level necessitated by the current experimental error, 
isospin-breaking (IB) corrections must be taken into account.

IB corrections for the $\pi\pi$ final
state, whose contributions dominate $\left[ a_\mu\right]_{had}^{LO}$,
were studied in Refs.~\cite{cen1,cen2} and, together with kinematic IB
corrections for the $4\pi$ contribution,
incorporated into the latest $\tau$-based $\left[ a_\mu\right]_{had}^{LO}$ 
analyses~\cite{dehz03,hockerichep04}. Even after
IB corrections, however, the high-precision CMD-2 $\pi\pi$ 
EM data~\cite{cmd203} lies $\sim 5-10\%$ below the IB-corrected $\tau$ data 
for $m_{\pi\pi}$ between $\sim 0.85$ and $\sim 1$ GeV~\cite{dehz03,ibfootnote}.
The corresponding determinations of $\left[ a_\mu\right]_{had}^{LO}$
differ by $\sim 2\sigma$, the $\tau$-based result lying 
higher and producing a SM $a_\mu$ prediction in significantly better 
agreement with 
experiment~\cite{dehz03,hmnt04,hockerichep04,otherdispersiveamu}.
Recent KLOE $e^+e^- \rightarrow \pi^+\pi^-$ cross-sections~\cite{kloe04}
yield an $\left[ a_\mu\right]_{had}^{LO}$ compatible
with CMD-2~\cite{hockerichep04}, though
the point-by-point agreement between the two data sets
is not entirely satisfactory~\cite{hockerichep04}.

In view of the unsettled experimental situation, we study 
sum rule constraints on weighted integrals
of the $I=1$ vector $\tau$ decay and EM spectral functions. 
The weights, $w(s)$, and upper integration limits, $s_0$, 
are chosen such that (i) each spectral integral
has a reliable and well-converged OPE representation,
and (ii) all relevant OPE input can be
obtained from sources independent of the low scale EM and $\tau$
data we seek to test. OPE uncertainties are minimized by working with $s_0$ 
and $w(s)$ for which the relevant OPE representation is dominated,
essentially entirely, by its $D=0$ component, and hence
determined, essentially entirely, by the single input 
parameter, $\alpha_s(M_Z)$, which can be taken from independent
high-scale studies. The $\tau$ decay based spectral integrals
will be shown to be well reproduced by
the corresponding OPE representations. Those based on
EM data, in contrast, will be shown to lie consistently below the 
corresponding OPE values, for positive $w(s)$, and to differ from 
them in their $s_0$ dependence. Both features are as expected 
if the EM spectral data is too low in the disputed region. 


\section{\label{sec:section2}The Sum Rule Constraints}

We study sum rule constraints on the EM spectral function,
$\rho_{EM}(s)$, and the sum of the spin $J=0$ and $1$ components
of the charged $I=1$ vector current spectral function,
$\rho_{V;ud}^{(0+1)}(s)\equiv \rho_{V;ud}^{(J=0)}(s) + \rho_{V;ud}^{(J=1)}(s)$.
The former is related to $R(s)$ by $\rho_{EM}(s)=R(s)/12\pi^2$, and
to the bare $e^+e^-\rightarrow hadrons$ cross-sections,
$\sigma_{bare}(s)$, by
\begin{equation}
\rho_{EM}(s)\, =\, s\, \sigma_{bare}(s)/16\pi^3\alpha_{EM} (0)^2\ .
\label{rhoemcrosssectionreln}\end{equation}
Defining $R_{V;ud}$ by 
$R_{V;ud}\equiv  \Gamma [\tau^- \rightarrow \nu_\tau
\, {\rm hadrons}_{V;ud}\, (\gamma)]/
\Gamma [\tau^- \rightarrow
\nu_\tau e^- {\bar \nu}_e (\gamma)]$
and $y_\tau\, \equiv\, s/m_\tau^2$, 
$\rho_{V;ud}^{(0+1)}(s)$ is related to $R_{V;ud}$ by
\begin{equation}
R_{V;ud}= 12\pi^2\vert V_{ud}\vert^2 S_{EW}\,
\int^{1}_0\, dy_\tau \,\left( 1-y_\tau\right)^2
\left[ \left( 1 + 2y_\tau\right)
\rho_{V;ud}^{(0+1)}(s) - 2y_\tau \rho_{V;ud}^{(0)}(s) \right]
\label{taukinspectral}
\end{equation}
where $V_{ud}$ is the 
flavor $ud$ CKM matrix element, and $S_{EW}$ is 
an electroweak correction~\cite{erler}. Contributions to
$\rho_{V;ud}^{(0)}(s)$ are of $O([m_d-m_u]^2)$, and 
hence numerically negligible, 
allowing $\rho_{V;ud}^{(0+1)}(s)$ to be determined from the experimental 
decay distribution. 

\subsection{\label{subsec:2a}Finite Energy Sum Rules}
For any correlator, $\Pi (s)$, with no kinematic singularities,
and any $w(s)$ analytic in $\vert s\vert <M$
with $M>s_0$, analyticity implies the finite
energy sum rule (FESR) relation
\begin{equation}
\int_0^{s_0}w(s)\, \rho(s)\, ds\, =\, -{\frac{1}{2\pi}}\oint_{\vert
s\vert =s_0}w(s)\, \Pi (s)\, ds\ ,
\label{basicfesr}
\end{equation}
where $\rho (s)$ is the spectral function of $\Pi (s)$. 
In QCD, for very large $s_0$ the OPE representation
can be employed on the RHS of Eq.~(\ref{basicfesr}).
As $s_0$ is decreased, this representation 
is expected to break down first near the timelike real $s$-axis~\cite{pqw}. 
A range of ``intermediate'' $s_0$ will thus exist for which the OPE, 
though unreliable for general $w(s)$, will remain valid for those
$w(s)$ satisfying $w(s=s_0)=0$. The corresponding FESR's are
called ``pinched'' or pFESR's.
For vector (V) and axial vector (A) correlators, and $w(s)=s^N$,
OPE breakdown (duality violation) is significant
at $s_0\sim$ a few GeV$^2$~\cite{kmfesr}. Polynomials $w(y)$ 
(with $y=s/s_0$) having even a single zero at $s=s_0$ ($y=1$), however, 
remove such violations for $s_0$ greater than 
$\sim 2$ GeV$^2$~\cite{kmfesr}, even for the
flavor $ud$ V-A correlator~\cite{vminusa}. 

In interpreting FESR results, one should bear in mind
that very strong correlations exist between spectral 
integrals corresponding to the same $w(y)$, but different $s_0$. 
Such correlations are particularly strong when $w(y)\geq 0$ over the 
relevant interval, $0<y\leq 1$, and even
more so when $w(y)$ is monotonically decreasing.
Similar strong correlations exist between spectral integrals corresponding 
to different $w(y)$, but fixed $s_0$. Correlations among the
corresponding OPE integrals are also very strong, especially when the OPE
is dominated, as below, by a single (in this case, $D=0$) contribution. 

We work, in what follows, with pinched polynomial weights, 
$w(y)=\sum_m c_my^m$. The pinching condition, $w(1)=0$ 
implies $\sum_m c_m=0$. 
For reasons explained below, $w(y)$ is further restricted to be
non-negative and monotonically decreasing on $0\leq y\leq 1$.
Since $s_0$ is the only scale entering the RHS of Eq.~(\ref{basicfesr}), 
it is obvious, on dimensional grounds, that integrated OPE contributions 
of dimension $D=2k+2$ scale as $1/s_0^k$, up to logarithms. For $D\geq 2$, 
such contributions are absent (up to corrections of $O(\alpha_s )$) 
when $c_{(D-2)/2}= 0$. The structure of the logarithmic integrals,
$\oint_{\vert s\vert =s_0}ds\, y^k\, \ell n(Q^2/\mu^2)/Q^D$, 
responsible for the $O(\alpha_s )$ corrections, is such that
cancellations inherent in the pinching condition $\sum_mc_m=0$
lead to strong numerical suppressions of these corrections.
$D\geq 8$ contributions are typically assumed
to be negligible, since the relevant condensate values 
are not known phenomenologically. The much stronger $s_0$ dependence 
of such high $D$ contributions allows this assumption to be tested
explicitly.

The reason for working with non-negative, monotonically decreasing $w(y)$ 
is that the EM spectral data for the $\pi^+\pi^-$ and $\pi^+\pi^-\pi^0\pi^0$
states, which dominate the EM-$\tau$ discrepancy, lie uniformly
below the IB-corrected $\tau$ data. Non-negativity of $w(y)$ then ensures 
that, for all $s_0$, the normalization of the $\tau$-based spectral 
integrals will be too high if it is the EM data which is
correct, while the normalization of the EM spectral integrals
will be too low if it is the $\tau$ data which is correct. 
Since the $y$ value for a given experimental bin decreases 
with increasing $s_0$, a monotonically decreasing $w(y)$ 
similarly ensures that the slope with respect to
$s_0$ of the $\tau$ spectral integrals will be too high if 
the EM data is correct, while the slope with respect to
$s_0$ of the EM spectral integrals will be too low if the
$\tau$ data is correct. The slope constraint is 
particularly useful because the slope of the corresponding OPE integrals 
is very tightly constrained, and only very weakly
dependent on the dominant OPE input parameter $\alpha_s(M_Z)$.

\subsection{\label{subsec:2b}OPE Input}
Re-writing the weighted pFESR OPE integrals of the relevant correlator, $\Pi$,
in terms of the Adler function $D(Q^2)\equiv -Q^2\, d\Pi (Q^2)/dQ^2$,
allows potentially large logs in the $D=0$ contribution
to be summed point-by-point along the integration contour.
This ``contour-improved'' (CIPT) prescription is known to 
improve the convergence of the integrated $D=0$ series~\cite{cipt}. The 
Adler function is given by
\begin{equation}
\left[ D(Q^2)\right]_{D=0}\, =\, C\, \sum_{k\geq 0}d^{(0)}_k \bar{a}^k\ ,
\label{dzeroadler}\end{equation}
where $\bar{a}=a(Q^2)=\alpha_s(Q^2)/\pi$ is the running coupling in the
$\overline{MS}$ scheme, and $C=1$, $2/3$ for the $\tau$, EM cases,
respectively. For $N_f=3$, $d^{(0)}_0=d^{(0)}_1=1$, 
$d^{(0)}_2=1.63982$ and $d^{(0)}_3=6.37101$~\cite{d0adler}. 
An estimate for $d^{(0)}_4$, $d^{(0)}_4=27\pm 16$~\cite{kataev}
also exists, based on methods which (i) work well for the 
coefficients of the $D=0$ series~\cite{bck02} and
(ii) produced, in advance of the actual calculation, 
an accurate prediction for the recently computed $O(a^3)$ $D=2$ 
coefficient of the $(J)=(0+1)$ V+A correlator sum~\cite{bck04}.

The leading $D=2$ contributions for the $\tau$ correlator
are $O(m_{u,d}^2)$ and numerically negligible. For the EM
correlator, up to tiny $O(m_{u,d}^2/m_s^2)$ 
corrections, the $D=2$ contributions are determined by
$\bar{a}$ and the running $\overline{MS}$ strange 
mass $\bar{m}_s$. At the scales employed here the integrated
$D=2$ contribution is small. The full expression
for $\left[ \Pi_{EM}(Q^2)\right]_{D=2}$ may be found in Ref.~\cite{bckem04}.
The $D=4$ terms in the OPE of the EM and $\tau$ correlators are 
determined by the RG invariant light quark, strange quark and gluon 
condensates, $\langle \bar{\ell}\ell\rangle_{RGI}$,
$\langle \bar{s}s\rangle_{RGI}$ and $\langle aG^2\rangle_{RGI}$,
up to numerically tiny $O(m_s^4)$ corrections. The expressions may
be found in Refs.~\cite{emd4ope,bnp}. The integrated $D=4$ contributions
are again small at the scales employed.
To reduce OPE uncertainties, we concentrate here
on weights for which the integrated leading $D=6$ contributions
are absent. More extensive studies will be reported elsewhere~\cite{kminprep}.
We assume throughout that contributions with $D\geq 8$ may be neglected,
but check this assumption for self-consistency, as discussed
above.

For $D=4$ input we 
use (i) $\langle aG^2\rangle =(0.009\pm 0.007)\ {\rm GeV}^4$
(from the recent re-analysis of charmonium sum rules~\cite{newgcond4})
and (ii) $\langle 2{m_\ell}\bar{\ell}\ell\rangle = -m_\pi^2f_\pi^2$
(the GMOR relation). $\langle m_s\bar{s}s\rangle_{RGI}$ then follows from
conventional ChPT quark mass ratios and the standard estimate
$\langle \bar{s}s\rangle_{RGI}/\langle \bar{\ell}\ell\rangle_{RGI} 
= 0.8\pm 0.2$.
For the $D=0,\ 2$ input, $\bar{a}$ and $\bar{m}_s$,
we employ exact solutions based on the
$4$-loop-truncated $\beta$ and $\gamma$ functions~\cite{betagamma},
with initial conditions
\begin{eqnarray}
&&m_s(2\ {\rm GeV})\, =\, 95\pm 20\ {\rm MeV}\label{msinput}\\
&&\alpha_s(M_Z)\, =\, 0.1200\pm 0.0020\label{alphasinput}.
\end{eqnarray}

Eq.~(\ref{msinput}) reflects the range of results
obtained in recent sum rule~\cite{bck04,mssumrule} and
$N_f=2$ and $2+1$ unquenched lattice studies~\cite{mslattice}.
The $N_f=5$ value in Eq.~(\ref{alphasinput}) is run down to the $N_f=3$ 
low-scale region using standard $4$-loop running and matching~\cite{cks97},
with the same matching scales as used in 
the recent EM sum rule studies of Refs.~\cite{hmnt04,hmnt03} (HMNT).
The input $\alpha_s(M_Z)$ in Eq.~(\ref{alphasinput}) 
differs from the PDG 2004 average for the following reasons.
First, the PDG average includes hadronic $\tau$ decay input, which must
be excluded if we wish to test the $\tau$ decay data. Second,
the PDG average is strongly affected by the quoted 
low, small-error determination from heavy quarkonium decay~\cite{pdg04rev}.
The Quarkonium Working Group, however, has (i) strongly criticized
the input to the low central value, (ii) argued that the quoted error
is underestimated by a factor of $3-5$~\cite{qwg04}, and (iii) concluded
the method is not competitive with extractions
based on perturbative treatments of high scale processes~\cite{footnote3}.
Eq.~(\ref{alphasinput}) is obtained by removing lower scale determinations,
including those based on heavy quarkonium and $\tau$ decay, from the PDG 
average~\cite{footnote4}.

\subsection{\label{subsec:2c}Spectral Input}
The ALEPH and CLEO $\tau$ decay distributions are in
good agreement. For definiteness, we use the ALEPH results,
for which the covariance matrix is publicly available. A small
global rescaling accounts for minor
changes in $B_e$, $B_\mu$, and the strange branching fraction, 
$B_s$, since the original ALEPH publication~\cite{aleph97}. 
PDG04~\cite{pdg04} values are used for $B_e$ and $B_\mu$, while
the updated $B_s$ value
incorporates (i) the new (2004) world averages for 
$B[\tau^-\rightarrow \nu_\tau K^-\pi^0]$ and
$B[\tau^-\rightarrow \nu_\tau K^-\pi^+\pi^-]$~\cite{opalstrange04}, 
(ii) the new (2005) CLEO results for the branching fractions 
of four-particle modes
with kaons~\cite{CLEOnewstrange05}, and (iii) the higher precision 
$K_{\mu 2}$ value for the $K$ pole contribution~\cite{pdg04}.
The long-distance EM corrections determined in Ref.~\cite{cen2} are also
applied to the dominant $\tau^-\rightarrow\pi^-\pi^0$ 
contribution~\cite{cen2}~\cite{thanksvc}.

Detailed discussions and assessments of the EM hadroproduction data base, as of
2002--2003, can be found in Refs.~\cite{dehz02,dehz03,hmnt04,whalley03}.
The exhaustive compilation of Ref.~\cite{whalley03} provides 
useful information on the treatment of radiative and VP corrections
for older experiments where such details 
are absent from the original publications. Detailed
assessments of the needed residual VP corrections 
are also contained in Refs.~\cite{dehz02,hmnt04}. 
These corrections are 
computed using the most recent version of 
F. Jegerlehner's code~\cite{jegerlehnervp}, generously 
provided by its author. 
We also employ the following new, high-precision
results, published
subsequent to the analyses of Refs.~\cite{dehz03,hmnt04}, 
and the compilation of Ref.~\cite{whalley03}:
(i) the final published
version of the SND $4\pi$ cross-sections~\cite{snd4pi03} 
(with systematic errors significantly reduced over those
of the earlier preprint version); (ii) the updated CMD-2 
$\pi^+\pi^-\pi^+\pi^-$~\cite{cmd24pi04} and
$\pi^0 \gamma$, $\eta \gamma$~\cite{cmd2pi0etagamma04} cross-sections;
and (iii) the BABAR $3\pi$~\cite{babar3pi} and
$\pi^+\pi^-\pi^+\pi^-$~\cite{babar4pi} cross-sections.
Since the $\pi\pi$ component of the EM-$\tau$ discrepancy
is driven by the CMD-2 data, with its very small $0.6\%$ systematic 
error~\cite{cmd203}, we employ only CMD-2 data in the CMD-2 region.
Where the existence of newer data permits, we
exclude older data for which systematic 
errors are incompletely stated, or absent, in the original publications,
and/or the residual radiative/VP corrections to be applied 
are unknown. Fortunately, data with missing systematic errors
for which no such replacement is possible
play only a small numerical role.
We assign a 
guess of $20\%$ for these errors in such cases.
For the small number of remaining older experiments where the
situation with regard to residual VP corrections is unclear,
we apply no VP correction. In all such cases, however,
(i) the corresponding contributions to the spectral integrals are small, and
(ii) the neglected VP corrections are, in any case,
much less than the quoted systematic errors.
The treatment of ``missing mode'' contributions, and the use
of isospin relations for a number of small contributions
where direct experimental determinations are absent, or have large errors,
follow the treatments discussed in detail in Refs.~\cite{dehz02,hmnt04}.
More details on the treatment of the EM data 
will be provided elsewhere~\cite{kminprep}.

\section{\label{sec:results}Analysis and Results}
For reasons discussed above, we concentrate on pFESR's involving
non-negative, monotonically decreasing $w(y)$. The only such degree $1$ 
weight is $w_1(y)=1-y$. Weights with a double zero at $y=1$, which
more strongly suppress OPE contributions from the vicinity of
the timelike axis, should be even safer from the point of view
of potential duality violation. A convenient set of
such ``doubly-pinched'' weights is the family,
$w_N(y)\, =\, 1-{\frac{N}{N-1}}\, y\, +\, {\frac{1}{N-1}}\, y^N,
\ N\geq 2$.
For a given $w_N$, the only non-$\alpha_s$-suppressed (``unsuppressed''
in what follows) $D>4$ OPE contribution is that with $D=2N+2$. This
contribution scales as $1/s_0^{N+1}$ 
relative to the leading integrated $D=0$ term. The strong $s_0$-dependence
allows the neglect of $D\geq 8$ contributions (for $w_{N\geq 3}$) 
to be tested for self-consistency. 
We have also studied pFESR's based on
a number of other weights. Since the results in all cases
point to the same conclusion, and OPE uncertainties are reduced for
weights having no unsuppressed $D=6$ contribution, we focus on the pFESR's
for two of the weights defined above, $w_1$ and $w_6$. Other results 
will be presented elsewhere~\cite{kminprep}.

Combined OPE errors for the various pFESR's 
are obtained by adding in quadrature uncertainties 
associated with the OPE input parameters
($D=4$ condensates, $m_s$, and $\alpha_s(M_Z)$) and 
the truncation/residual scale dependence 
of the integrated $D=0,\ 2$ series. The latter are
estimated to be {\it twice} the magnitude of the last term
in the corresponding truncated series. The resulting OPE error estimate
is somewhat more conservative than that employed by HMNT.

\begin{figure}
\unitlength1cm
\caption{EM OPE and spectral integrals for the weight $w_1$} 
\begin{minipage}[t]{8.2cm}
\begin{picture}(8.1,8.1)
\epsfig{figure=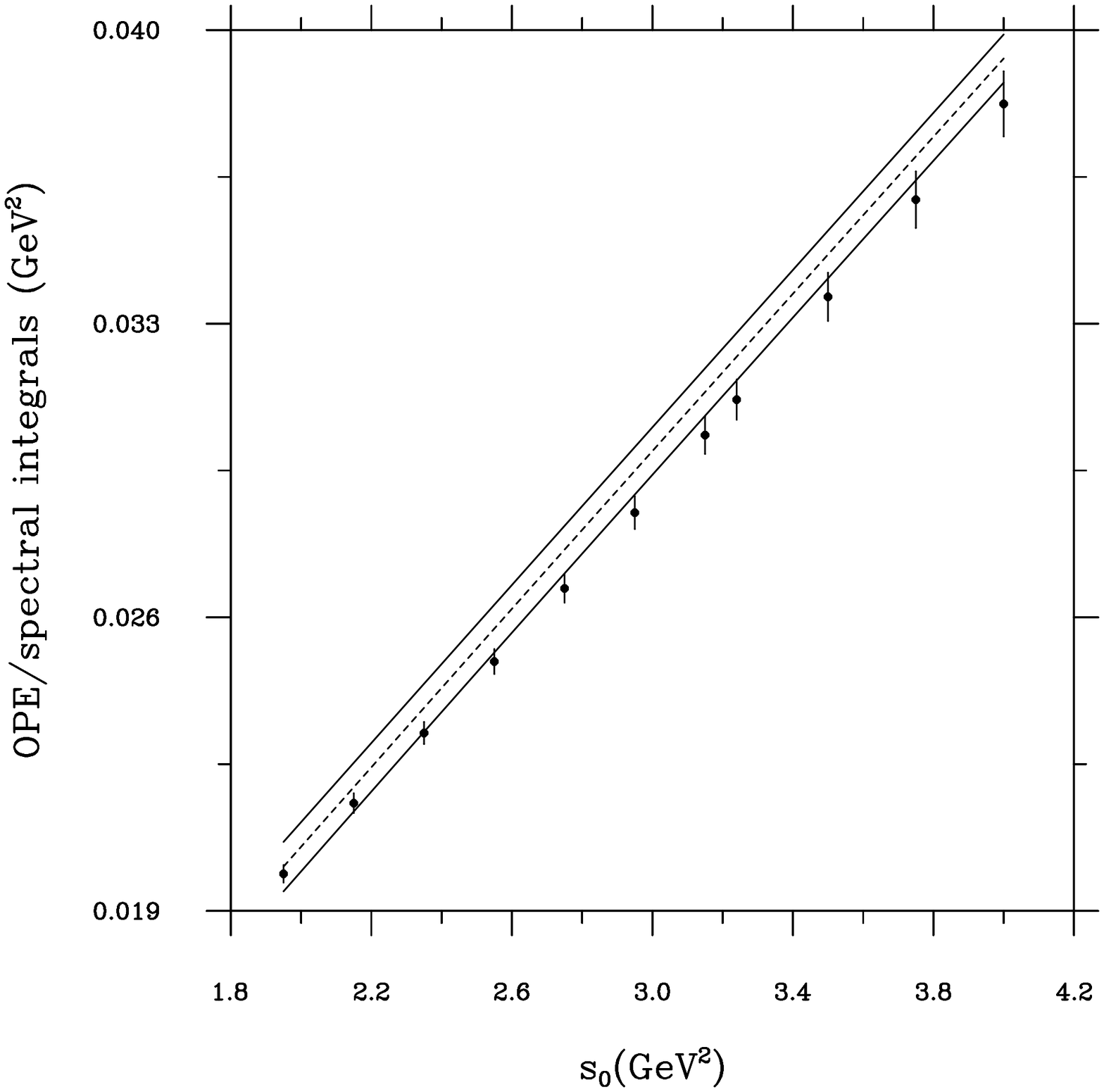,height=8.0cm,width=8.0cm}
\end{picture}
\end{minipage}
\label{fig1a}\end{figure}

\begin{figure}
\unitlength1cm
\caption{EM OPE and spectral integrals for the weight $w_6$}
\begin{minipage}[t]{8.2cm}
\begin{picture}(8.1,8.1)
\epsfig{figure=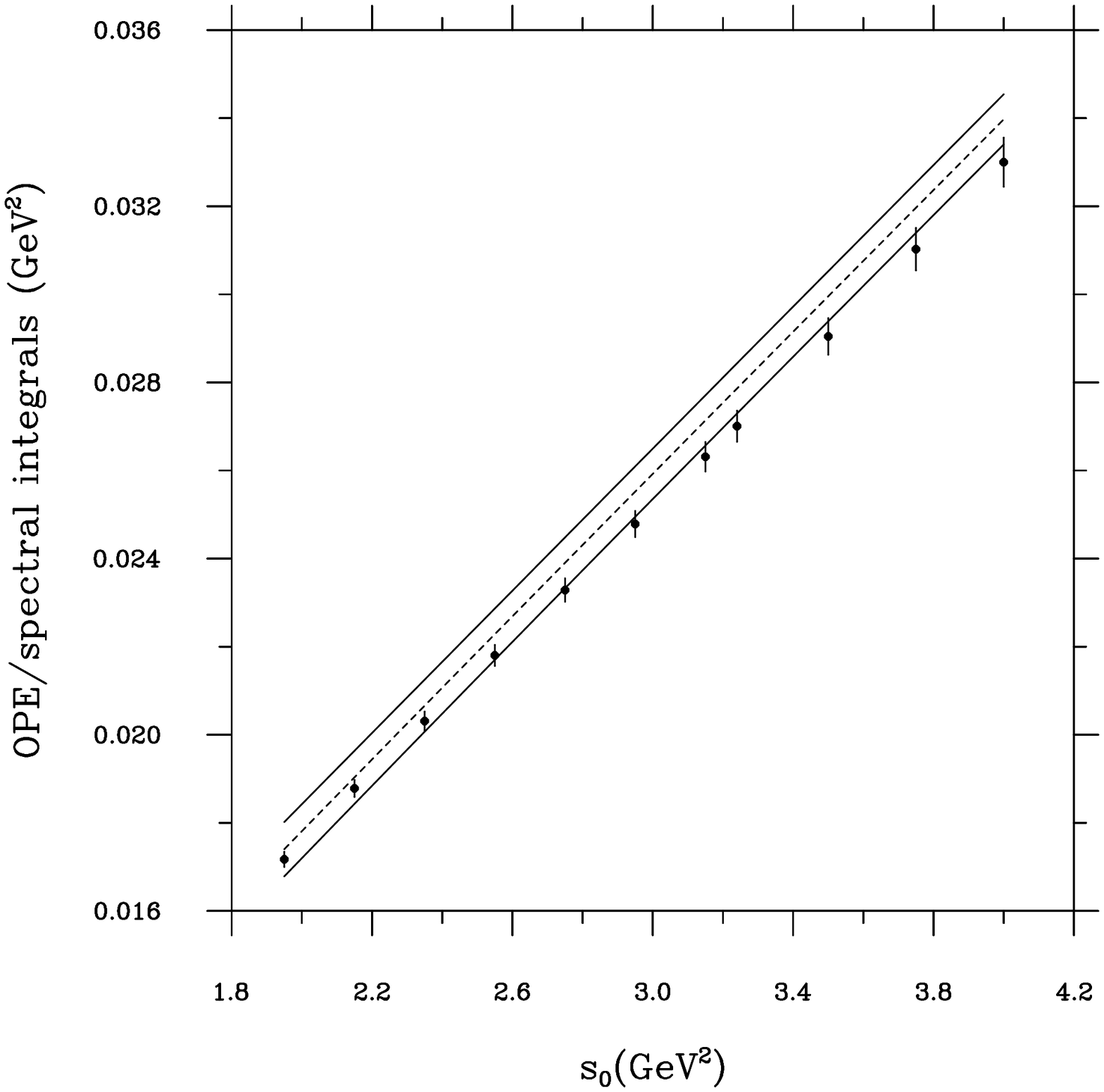,height=8.0cm,width=8.0cm}
\end{picture}
\end{minipage}
\label{fig1c}\end{figure}

\begin{figure}
\unitlength1cm
\caption{$\tau$ OPE and spectral integrals for the weight $w_1$}
\begin{minipage}[t]{8.2cm}
\begin{picture}(8.1,8.1)
\epsfig{figure=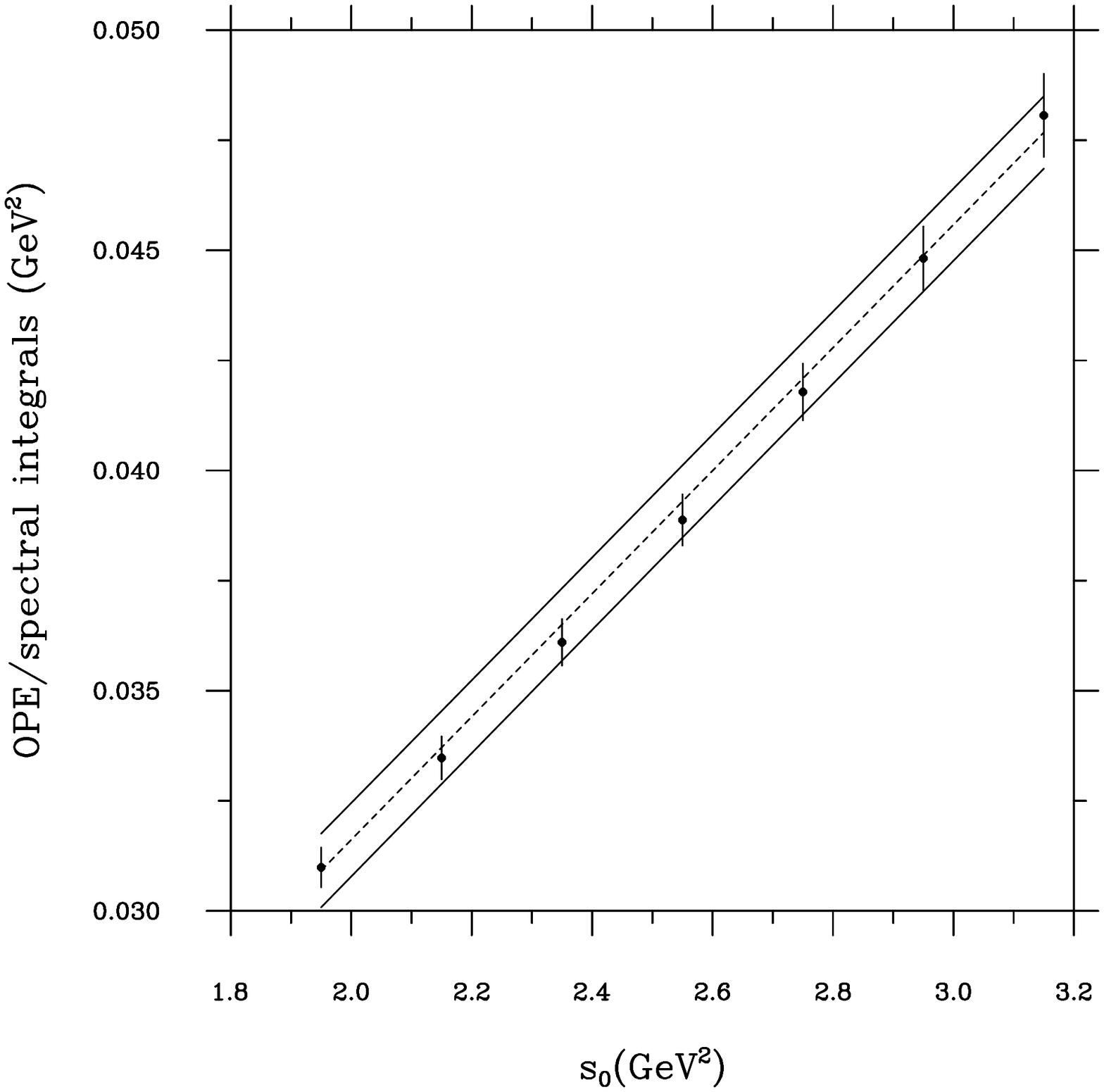,height=8.0cm,width=8.0cm}
\end{picture}
\end{minipage}
\label{fig2a}\end{figure}

\begin{figure}
\unitlength1cm
\caption{$\tau$ OPE and spectral integrals for the weight $w_6$}
\begin{minipage}[t]{8.2cm}
\begin{picture}(8.1,8.1)
\epsfig{figure=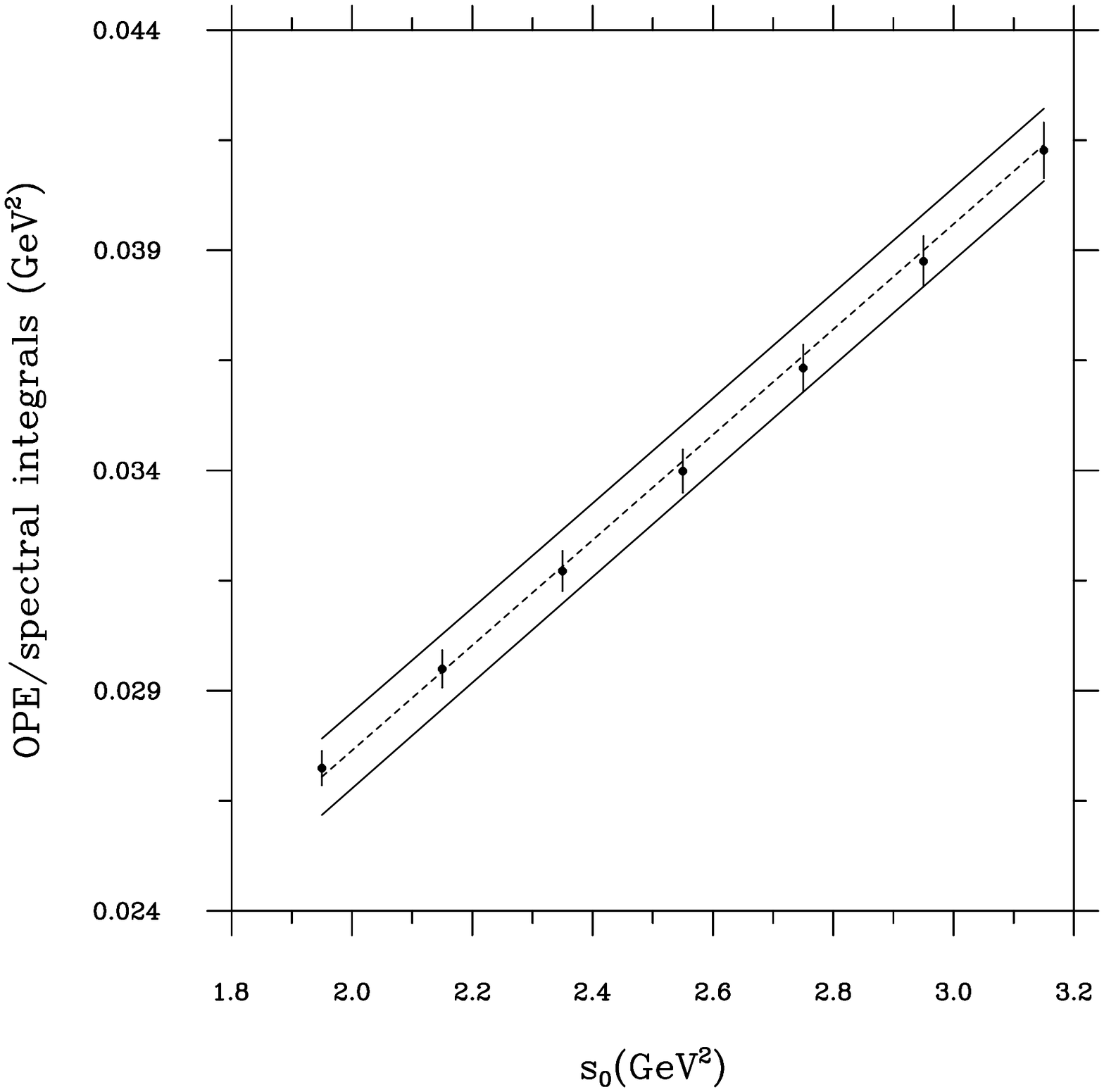,height=8.0cm,width=8.0cm}
\end{picture}
\end{minipage}
\label{fig2c}\end{figure}

Results for the EM case are presented in Figures~\ref{fig1a}, ~\ref{fig1c},
those for the $\tau$ case in Figures~\ref{fig2a}, ~\ref{fig2c}.
The dashed lines represent the central OPE results, the solid lines the upper 
and lower edges of the OPE error bands. Because of the
strong correlations, the OPE band
is better thought of as a bundle of allowed parallel
lines than as a generally allowed region. We see that, in the region 
$2\ {\rm GeV}^2<s_0<m_\tau^2$, both the magnitude and slope of the integrals
over the $\tau$ decay distribution are in good agreement with OPE expectations.
In contrast, the EM spectral integrals are consistently low
relative to OPE expectations (particularly for $s_0$ greater than 
$\sim 2.5\ {\rm GeV}^2$) and have slopes with respect
to $s_0$ significantly lower than those of the OPE curves. 
Both features of the EM results are as expected if the $\tau$ data 
is correct, and hence the EM data low, in the disputed region.

The implications of the normalizations of the EM and $\tau$ 
spectral integrals can be quantified by working out the
$\alpha_s(M_Z)$ required to match the OPE and spectral sides of a given 
pFESR. The reliability of the OPE, and hence of the extraction of 
$\alpha_s(M_Z)$, is optimized by choosing
$s_0$ as large as possible -- in the case of hadronic $\tau$ decay, 
$s_0=m_\tau^2$. Results corresponding to central input for the
small $D=2,4$ OPE contributions and the $s_0=m_\tau^2$ values
of the spectral integrals are shown in Table~\ref{table1}
for both the EM and $\tau$ cases.
The agreement between the $\tau$-decay and independent high-scale
determinations is excellent~\cite{taunormfootnote}. In contrast, the EM data
corresponds to $\alpha_s(M_Z)$ $\sim 2\sigma$
lower than the high-scale determination. The $y^6$ term of
$w_6(y)$, in principle, produces an unsuppressed $D=14$
OPE contribution scaling as $1/s_0^6$ ($1/s_0^7$ relative to
the leading $D=0$ term). Such a contribution, if present, would 
contaminate the extraction of $\alpha_s(M_Z)$. 
There is, however, no evidence for such a contribution,
at a level which would impact our analysis,
in the $s_0$ dependence of either the EM or $\tau$ 
$w_6$-weighted spectral integrals~\cite{highdfootnote}.
The excellent agreement between the $\alpha_s(M_Z)$
extracted using different doubly-pinched weights,
with potential unsuppressed $D>6$ contributions of different dimension,
provides further evidence in support of the absence of such
$D>6$ contributions~\cite{normfootnote,realephd8footnote}.

\begin{table}
\caption{\label{table1}The values of $\alpha_s(M_Z)$ obtained
by fitting to the experimental EM and $\tau$ spectral integrals
for $s_0=m_\tau^2$, with central values for the $D=2,4$ OPE input}
\vskip .1in
\begin{tabular}{ccc}
\hline
Weight&\qquad$\left[ \alpha_s(M_Z)\right]_{EM}$\qquad&
\qquad$\left[ \alpha_s(M_Z)\right]_{\tau}$\qquad\\
\hline
$w_1$&$0.1138^{+0.0030}_{-0.0035}$&$0.1212^{+0.0027}_{-0.0032}$\\
$w_6$&$0.1150^{+0.0022}_{-0.0026}$&$0.1195^{+0.0020}_{-0.0022}$\\
\hline
\end{tabular}
\end{table}

To quantify the disagreement between the EM OPE and
experimental slope values, we work out the correlated
errors for the slopes with respect to $s_0$ of the OPE and spectral integrals. 
The correlations are such that
the uncertainty on the OPE side is rather small. In particular,
the slope is quite insensitive to $\alpha_s(M_Z)$. These points are
illustrated in Table~\ref{table2}, which shows the
spectral integral and OPE slope values for the EM $w_1$ and $w_6$ 
pFESR's. The OPE entries labelled ``indep'' are those obtained using
the independent, high-scale fit value for $\alpha_s(M_Z)$.
Those labelled ``fit'', in contrast, correspond to the
$\alpha_s(M_Z)$ values obtained by fitting to the relevant
$s_0=m_\tau^2$ spectral integrals, as given in Table~\ref{table1}.
We see that, even if one were willing to tolerate the lower central 
$\alpha_s(M_Z)$ values implied by the EM spectral integrals,
such a lowering of $\alpha_s(M_Z)$ would have negligible impact on the OPE vs.
spectral integral slope discrepancy problem. 

\begin{table}[ht]
\caption{\label{table2}Slopes wrt $s_0$ of the EM OPE and spectral integrals}
\vskip .1in
{\begin{tabular}{cccc}
\hline
Weight&$S_{exp}$&$\alpha_s(M_Z)$&$S_{OPE}$\\
\hline
$w_1$&$.00872\pm .00026$&indep&$.00943\pm .00008$\\
&&fit&$.00934\pm .00008$\\
\hline
$w_6$&$.00762\pm .00017$&indep&$.00811\pm .00009$\\
&&fit&$.00805\pm .00009$\\
\hline
\end{tabular}}
\end{table}

\section{\label{sec:conclusions}Discussion and Conclusions}
We have shown that weighted spectral integrals
constructed using $I=1$ hadronic $\tau$ decay 
data are in good agreement with OPE expectations, while 
those involving EM data (i) require a value of $\alpha_s(M_Z)$ 
$\sim 2\sigma$  below that given by high-scale determinations
and (ii) correspond to a slope with respect to $s_0$ in
$\sim 2.5\sigma$ disagreement with the OPE prediction. 
The slope problem, moreover,
{\it cannot} be cured simply by adopting the lower $\alpha_s(M_Z)$ values,
shown in Table~\ref{table1}, which would bring the normalization
of the OPE and spectral integrals into agreement for 
$s_0\simeq m_\tau^2$. The insensitivity of the slope to $\alpha_s(M_Z)$ also
means that the agreement between the OPE expectation and the observed 
slope for the $\tau$ decay spectral integrals represents a 
non-trivial test of the $\tau$ data.

One possibility is that the problems with the EM sum rules 
might be attributable to the presence of residual duality violation at
the intermediate scales studied here; the success of the OPE in 
predicting both the slope and magnitude of the $\tau$-decay-based spectral 
integrals over the whole of the region $2\ {\rm GeV}^2< s_0< m_\tau^2$,
however, renders such an explanation highly implausible.
The results thus point to the reliability of the $\tau$ data,
and to the likelihood of either (i) a problem with 
the experimental EM spectral distribution, or (ii) the presence
of as-yet-unidentified non-one-photon physics contributions
in the experimental EM cross-sections. This in turn suggests that
$a_\mu$ determinations which incorporate $\tau$ decay data
are to be favored over those employing EM spectral data only.

While the disagreement between the EM and high-scale determinations
of $\alpha_s(M_Z)$ is only $\sim 2 \sigma$, even with
significantly lower high-scale input, e.g., the 2002 PDG average 
$\alpha_s(M_Z)=0.1172\pm 0.0020$ used by HMNT, the EM normalization,
and even more so the EM slope, would still require,
on average, upward fluctuations in $\rho_{EM}(s)$.
Since $K(s)/s>0$, such fluctuations 
would typically also increase $\left[ a_\mu\right]_{had}^{LO}$. 
Thus, even ignoring the slope problem and assessing the EM data 
as moderately consistent, within errors, with the OPE constraints, 
the fact that the spectral integrals lie consistently below the
corresponding OPE constraint values, for any sensible input
$\alpha_s(M_Z)$, points to the likelihood
of a $\left[ a_\mu\right]_{had}^{LO}$ value 
higher than the current central EM-data-based value. 

Two further points are of relevance to assessing the implications
of our results for the value of $\left[ a_\mu\right]_{had}^{LO}$.
First, we find that,
replacing the EM $\pi^+\pi^-$, $\pi^+\pi^-\pi^0\pi^0$ and
$\pi^+\pi^-\pi^+\pi^-$ data with the equivalent $\tau$ data 
resolves completely both the normalization and slope problems for the 
resulting modified ``EM'' spectral integrals~\cite{tautoemfootnote}. 
Second, it is readily demonstrated that the pFESR's employed 
are sensitive to, not just the discrepancies in the $4\pi$ region,
but also those in the $2\pi$ region.
(This is relevant since the $\left[ a_\mu\right]_{had}^{LO}$ integral
is dominated by the $2\pi$ spectral contribution.)
In fact, for the $w_1$ pFESR, the shift in
the EM spectral integral associated with the modification of the $2\pi$ 
part of the EM spectral function represents
$82\%$ of the full shift at $s_0=2\ {\rm GeV}^2$
and $32\%$ at $s_0=m_\tau^2$. The corresponding
figures for the $w_6$ pFESR are $87\%$ at
$s_0=2\ {\rm GeV}^2$ and $45\%$ at
$s_0=m_\tau^2$. Thus, even though the pFESR's employed are
relatively more sensitive to the $4\pi$ spectral contributions than
is the $\left[ a_\mu\right]_{had}^{LO}$ integral, a
clear sensitivity to the $2\pi$ component remains, making the
constraints associated with these pFESR's highly relevant to the
$\left[ a_\mu\right]_{had}^{LO}$ problem.

In conclusion, all the sum rule tests performed favor the
reliability of the $\tau$ decay data, and point to problems
with the EM data. We conclude that, at present, determinations
of $\left[ a_\mu\right]_{had}^{LO}$ employing IB-corrected $\tau$ decay data
are more reliable than those based on EM data alone, and
hence that there is no clear sign of a discrepancy between
the current experimental value for $a_\mu$ and the
SM prediction.
\vskip .15in\noindent
{\it NOTE ADDED: Subsequent to the submission of this paper, new results for
the $e^+e^-\rightarrow\pi^+\pi^-$ cross-sections were released by
the SND Collaboration~\cite{snd05}. As would be expected from
the sum rule results above, the SND cross-sections are compatible
with the $\tau\rightarrow\nu_\tau\pi\pi$ data, but in significant disagreement
with the KLOE $\pi\pi$ data, in the disputed region.}

\begin{acknowledgments}
Thanks to Daisuke Nomura and Thomas Teubner
for additional details on
the calculations of Refs.~\cite{hmnt03,hmnt04}; 
Simon Eidelman for useful input on the EM spectral data;
Thomas Teubner and Mikhail Achasov for clarifying the treatment
of VP effects in the SND EM data; Fred Jegerlehner for providing
a copy of his code for computing VP corrections; and Vincenzo
Cirigliano for providing explicit numerical results for
the long-distance EM corrections to the $\tau$ $\pi\pi$ 
distribution. 
The hospitality of the CSSM, University of
Adelaide, and ongoing support of the Natural Sciences and Engineering 
Council of Canada are also gratefully acknowledged.
\end{acknowledgments}


\vfill\eject
\hfill\vskip .1in
\vfill\eject
\hfill\vskip .1in


\begin{thebibliography}{99}
\bibitem{ibfootnote}In Ref.~\cite{kmcewib05}, it was pointed out that
some model dependence exists in the determination of the $\rho$-$\omega$
``mixing'' component of the IB correction, and also that an IB isoscalar
current contribution to the broad $\rho$ distribution might be present 
in the experimental data. Neither of these features was taken into account
in previous treatments of the IB correction. The size of these effects,
however, is far too small to account for the discrepancy between the
CMD2 and IB-corrected $\tau$ data.
\bibitem{footnote3}The low $Q^2$ region, where scaling violations 
are largest, will also dominate determinations based on scaling violations
in DIS.
\bibitem{taunormfootnote}Part of the error quoted for 
$\alpha_s(M_Z)$ in the $\tau$ case is that associated with the $\sim 0.7\%$
normalization uncertainty on the $\tau$-based spectral
distribution. This uncertainty contributes
$\pm 0.0010$ to the uncertainty on $\alpha_s(M_Z)$, so lower
central values of $\alpha_s(M_Z)$ are easily accommodated
within the normalization uncertainty. The values quoted
in the table would be $0.0005$ higher if
the long-distance EM corrections of Ref.~\cite{cen2} were neglected.
\bibitem{highdfootnote}An extremely conservative upper bound on the
impact of such a term can be obtained by adding a $D=14$ contribution to
the OPE side and performing a fit for its strength,
ignoring the very strong correlations between $w_6$-weighted 
integrals at different $s_0$. In the $\tau$ case,
the resulting $D=14$ contribution is $<5\%$
of the $D=0$ contribution at $s_0=2\ {\rm GeV}^2$, 
and hence $<0.2\%$ at $s_0=m_\tau^2$. Incorporating
such a contribution would shift the extracted value of
$\alpha_s(M_Z)$ by $<0.0003$. Accounting for the strong
correlations would lead to an even smaller shift. 
Note also that it is impossible to attribute the EM
slope and normalization problems to
neglected higher $D$ contributions: a $D=14$
term with strength sufficient to bring the $w_6$-weighted EM OPE and spectral
integrals into agreement at $s_0\simeq 4\ {\rm GeV}^2$ would
be $\sim 4$ times the corresponding $D=0$ contribution at 
$s_0=2\ {\rm GeV}^2$, and produce
a disastrously bad match between the OPE and spectral integral curves.
\bibitem{normfootnote}The $\alpha_s(M_Z)$ values
extracted using the $w_3,\, w_4$, and $w_5$ sum rules are,
for the EM case, $0.1152^{+0.0019}_{-0.0021}$, $0.1154^{+0.0020}_{-0.0023}$,
and $0.1152^{+0.0022}_{-0.0024}$, respectively, and, for the $\tau$ case,
$0.1189^{+0.0018}_{-0.0021}$, $0.1193^{+0.0019}_{-0.0022}$,
and $0.1194^{+0.0020}_{-0.0022}$, respectively. In both cases,
the results are in excellent agreement with those obtained 
using the corresponding $w_6$ sum rule.
\bibitem{tautoemfootnote}The normalizations of the modified EM
integrals are now in full agreement with the high-scale-input OPE
constraint bands, for all $s_0>2\ {\rm GeV}^2$. The 
modified spectral integral slopes are also in good agreement 
with OPE expectations (the modified central values are $0.00947$ for
the $w_1$ and $0.00807$ for the $w_6$ pFESR, to be
compared with the OPE predictions $0.00943\pm 0.0008$ 
and $0.00811\pm 0.00009$ already reported in Table~\ref{table2}.
\end{thebibliography}
\end{document}